\documentstyle[eqsecnum,aps]{revtex}

\input epsf.tex

\def\sl#1{\rlap{\hbox{$\mskip 1 mu /$}}#1}      
\def\bo{\raisebox{-.4ex}{\large$\Box$}}                 

\def\a{\alpha}
\def\b{\beta}
\def\c{\chi}
\def\d{\delta}
\def\e{\epsilon}                
\def\f{\phi}                    
\def\g{\gamma}

\def\j{\psi}
\def\k{\kappa}
\def\l{\lambda}
\def\m{\mu}
\def\n{\nu}

\def\p{\pi}                     
\def\th{\theta}                  
\def\s{\sigma}                  

\def\x{\xi}
\def\z{\zeta}
\def\D{\Delta}
\def\F{\Phi}
\def\G{\Gamma}

\def\L{\Lambda}

\def\P{\Pi}

\def\cd{{\cal D}}
\def\cf{{\cal F}}
\def\cg{{\cal G}}
\def\ch{{\cal H}}

\def\cl{{\cal L}}
\def\co{{\cal O}}
\def\cp{{\cal P}}

\def\svev#1{\left\langle #1\right\rangle}       

\def\beq{\begin{equation}}
\def\eeq{\end{equation}}
\def\bqry{\begin{eqnarray}}
\def\eqry{\end{eqnarray}}
\def\NON{\nonumber\\}
\def\seeq#1{eq.~(\ref{#1})}
\def\seEq#1{Eq.~(\ref{#1})}
\def\seeqs#1{eqs.~(\ref{#1})}

\def\seneq#1{~(\ref{#1})}

\def\tr{{\rm tr}\,}

\def\det{{\rm det}\,}
\def\rhs{\mbox{r.h.s.} }

\def\eg{\mbox{\it e.g.} }

\def\hc{\mbox{h.c.} }
\def\half{{1\over 2}}

\def\rf{ref.~\cite}

\def\CMP#1{Comm. Math. Phys. {\bf #1}}

\def\JMP#1{Jour. Math. Phys. {\bf #1}}
\def\NPB#1{Nucl. Phys. {\bf B#1}}
\def\NPBP#1{Nucl. Phys. (Proc. Suppl.) {\bf B#1}}
\def\PLB#1{Phys. Lett. {\bf B#1}}
\def\PRD#1{Phys. Rev. {\bf D#1}}

\def\PRL#1{Phys. Rev. Lett. {\bf #1}}

\def\PTP#1{Prog. Theor. Phys. {\bf #1}}

\def\tk{\tilde\k}
\def\AM{{AM}}

\def\half{{\scriptstyle {1\over 2}}}

\begin{document}
\draft
\preprint{TAUP--2306--95}
\twocolumn[\hsize\textwidth\columnwidth\hsize\csname%
@twocolumnfalse\endcsname

\title{
The Standard Model from a New Phase Transition on the Lattice}
\author{Yigal Shamir\cite{supp}}
\address{ School of Physics and Astronomy,
Beverly and Raymond Sackler Faculty of Exact Sciences\\
Tel-Aviv University, Ramat Aviv 69978, ISRAEL}
\date{\today}
\maketitle

\begin{abstract}
Several years ago it was conjectured in the so-called 
Roma Approach~\cite{roma}, that gauge fixing is an essential ingredient
in the lattice formulation of chiral gauge theories.
In this paper we discuss in detail how the gauge-fixing approach may
be realized. As in the usual (gauge invariant) lattice formulation, 
the continuum limit corresponds to a gaussian 
fixed point, that now controls both the transversal and the longitudinal
modes of the gauge field. A key role is played by a new phase 
transition separating a conventional Higgs or Higgs-confinement phase, 
from a phase with broken rotational invariance. 
In the continuum limit we expect to find a scaling region, 
where the lattice correlators reproduce the euclidean 
correlation functions of the target (chiral) gauge theory,
in the corresponding continuum gauge. 
\end{abstract}
\pacs{11.15.Ha 12.15.-y 11.30.Rd}

]

\narrowtext



\section{Introduction}

  The great difficulty in constructing chiral gauge theories, 
such as the Standard Model, using the lattice regularization 
is related to the {\it doubling problem}~\cite{kw,KS,NN,ys}.
In lattice QCD, species doubling occurs when the discretized
fermion action has an unwanted symmetry 
that should be anomalous in the continuum. When one uses Wilson fermions,
the Wilson term eliminates the doublers, at the price of
breaking all the axial symmetries explicitly.
In the continuum limit, one expects to recover the axial symmetries, 
except the anomalous U(1), by tuning the fermion hopping parameter 
to a critical value. 

  In the lattice discretization of a chiral gauge theory, 
one has to account for the fact that a Weyl
fermion in a complex representation contributes 
to the gauge anomaly. This means that a lattice action for a single 
chiral fermion cannot be gauge invariant. We will assume below that,
as in the continuum, the lattice fermion action involves a sum over 
different complex representations, whose total gauge anomaly is zero. 
The question is to what extent the violations of gauge invariance, 
coming from the individual representations, cancel each other. 

  Consider the {\it regularized} effective action 
obtained by integrating out an anomaly-free set of chiral fermions.
In the continuum, one can use Dimensional Regularization to define the
effective action for smooth gauge fields 
that vanish rapidly at infinity. Using the freedom to add 
local counter-terms, the violations of gauge invariance 
are proportional to the dimensionless parameter $\e=d-4$, 
and so they vanish in the limit $\e \to 0$. This extends to
topologically non-trivial background fields, using for example 
the $\z$-function regularization~\cite{zeta}.

  On the lattice one encounters a fundamentally different situation.
The lattice spacing is a dimensionful parameter that plays a dual role.
First, it provide a UV cutoff, by replacing the infinite range 
of momentum integrals with an integration over the periodic Brillouin zone.
In addition, the lattice spacing enters the (multi-valued) mapping from
the compact link variables $U_{x,\m}$, to the Lie-algebra valued $A_\m$ field. 
These differences in the global structure imply that
a generic {\it lattice} gauge transformation, considered as a mapping
that acts on the Fourier space of the lattice $A_\m$-field, 
is qualitatively different from the corresponding mapping 
defined by a {\it continuum} gauge transformation.
The result is that the lattice effective action suffers from generic
violations of gauge invariance which are not controlled by any small parameter. 
We know of no method that ensures
the smallness of these violations on the entire lattice gauge orbit,
at the price of tuning any finite number of parameters
(see Sect.~\ref{sect4c} or the review article~\cite{rev} for more details).

  In the so-called Roma Approach~\cite{roma}, it was conjectured that
gauge fixing is a crucial ingredient in the lattice formulation
of chiral gauge theories. A gauge fixing action should assign
a bigger Boltzmann weight to a smooth gauge field,
relative to a rough field that belongs to the same gauge orbit.
This should reduce lattice-artefact violations 
of gauge invariance, because the latter are associated with
the roughness of the lattice gauge field. 

  In spite of this promising picture, the gauge-fixing approach has 
remained elusive. A naive discretization of the Lorentz 
gauge-fixing action  leads to a lattice action that has  a dense set of 
{\it lattice} Gribov copies with no continuum counterparts. 
These lattice artefact Gribov copies exist even for the classical vacuum. 
(Remarkably, the proliferation of Gribov copies
on the lattice resembles the fermion doubling problem in a number of ways.) 
As a result, the Boltzmann weight of too many rough lattice configurations
is not suppressed.

  In this paper we construct a lattice gauge-fixing action that
accommodates this problem. (See \rf{pre} for a preliminary version
of this work.) The gauge-fixing action is associated 
with a new generic phase diagram. 
We argue that, in this phase diagram, there is a gaussian critical point 
that belongs to the universality class of a gauge-fixed 
continuum theory. In comparison with the gauge-invariant lattice 
definition of QCD, the weak coupling limit here controls not
only the transversal modes of the gauge field, but also the longitudinal ones.

  As discussed above, gauge invariance of the target continuum theory,
as well as the residual BRST invariance, are both explicitly broken 
on the lattice. By tuning a finite number of counter-terms,
one hopes to recover BRST invariance in the
continuum limit, provided the fermion spectrum is anomaly-free~\cite{roma}.
The BRST identity that requires the vanishing of the renormalized 
gauge boson mass, $m_r^2=0$, plays a key role. Since the regularization 
is not gauge invariant, a mass counter-term has to be introduced, 
and its parameter needs to be tuned, in order to enforce
this BRST identity. Usually, a negative renormalized mass-squared 
indicates spontaneous symmetry breaking. 
Here we encounter a new feature, namely, the {\it gauge field} condenses 
if its mass-squared parameter becomes too negative. This implies that the 
new critical point is located on the boundary between a conventional phase, 
which is invariant under lattice rotations, and a new phase where 
the lattice rotation symmetry is broken spontaneously
by the VEV of the gauge field.

  The construction of the gauge-fixing action is presented in Sect.~\ref{sect2}.
The main results are (a) the gauge-fixing action has a unique absolute
minimum,  $U_{x,\m}=I$, 
and (b) perturbation theory around this minimum is 
manifestly renormalizable. In Sect.~\ref{sect3} we discuss a simple chiral 
fermion action. The validity of perturbation theory
implies the onset of a scaling behaviour in the weak-coupling limit.
Up to the regularization-dependent counter-terms,
the continuum lagrangian that controls the scaling behaviour 
can be read off from the marginal and relevant terms of the
lattice action. The scaling region should therefore faithfully
reproduce the correlation functions of the target chiral gauge theory,
in the corresponding gauge.

  Without the new gauge-fixing action, the fermion action of Sect.~\ref{sect3}
{\it does not} lead to a chiral gauge theory in the continuum limit. 
The longitudinal modes fluctuate strongly, and their non-perturbative 
dynamics ultimately renders the fermion spectrum vector-like. 
(This applies to many other chiral fermion proposals, see \rf{rev,revDP}.)  
In the second part of this paper, we examine the dynamics of the lattice
longitudinal modes from a broader point of view. We explain
the problems created by this dynamics, and how they may be solved
within the present approach.

  In Sect.~\ref{sect4} we discuss the lattice effective action, and the role of
lattice artefact Gribov copies.
In Sect.~\ref{sect5} we discuss the complete phase diagram in the limit of a
vanishing gauge coupling. In Sect.~\ref{sect6} we explain 
how our approach evades the No-Go theorems. 
Several open questions are discussed in Sect.~\ref{sect7}, and our
conclusions are offered in Sect.~\ref{sect8}.

\newpage
\section{Constructing the lattice gauge-fixing action}
\label{sect2}

\subsection{The phase transition associated with\protect\\ 
a critical vector boson}
\label{sect2a}

This section is devoted to a step-by-step construction of
the gauge-fixing action. As discussed in the introduction, 
the lattice-regularized theory has no symmetry that protects 
the masslessness of the vector bosons. Therefore, in the relevant
part of the phase diagram, the lattice vector field is generically not
critical. Now, according to the standard lore, the correlation length 
should diverge close to a continuous phase transition 
associated with the condensation a Bose field.
In this paper we wish to apply this to the lattice vector field.

  As a preliminary requirement for a continuous transition, one needs
a higher-power term, that stabilizes the classical potential
$V_{\rm cl}(A_\m)$ when the coefficient
of the quadratic term changes sign. In a weakly-coupled theory,
the actual location of the transition should be close to its
tree-level value, and near the transition
one expects the onset of a scaling behaviour governed by 
renormalized (continuum) perturbation theory. 

  Our aim is to achieve criticality of the
lattice vector field, very much like the way this is done in
the familiar $\F^4$ theory. However, going from a spin-0 to
a spin-1 field presents new difficulties.
The lattice theory is formulated in terms of the link variables $U_{x,\m}$,
which are group-valued parallel transporters. On the other hand, 
renormalized perturbation theory,
that governs the scaling region, is more naturally formulated in terms 
of the Lie-algebra valued $A_\m$ field. Thus, it takes some trial and error
to find the $U_\m$-dependent action that best suits our purpose.

  Another complication arises because not every 
renormalizable vector theory is unitary. A unitary, physical Hilbert space 
exists if and only if the vector theory is actually an 
anomaly-free gauge theory in a gauge fixed form.
This requires us to choose the lattice action, such that the  marginal gauge
symmetry breaking terms in the tree-level vector lagrangian have the form
\beq
  {1\over 2 \x_0}\, (\mbox{gauge condition})^2 \,.
\label{gcond}
\eeq
An appropriate Faddeev-Popov ghost action will be necessary too.

\subsection{A higher-derivative Higgs action}
\label{sect2b}

Our starting point is the lattice action
\beq
  S = S_G(U) + S_H(\f,U)\,.
\label{s1}
\eeq 
Here $S_G(U)$ is the usual plaquette action. The Higgs action is
\beq
  S_H = \tr \sum \left(
         -\k\, \f^\dagger\bo(U)\f + \tk\, \f^\dagger \bo^2(U) \f \right)\,,
\label{sh}
\eeq
where 
\beq
  \bo_{xy}(U) = \sum_\m ( \d_{x+\hat\m,y} U_{x,\m} 
  + \d_{x-\hat\m,y} U^\dagger_{y,\m}) - 8 \d_{x,y} \,,
\label{bo}
\eeq
is the standard nearest-neighbour covariant laplacian.
The lattice spacing $a$ is equal to one. Both $U_{x,\m}$ and $\f_x$
take values in a Lie group $G$. The first term on the \rhs of 
\seeq{sh} is a conventional lattice Higgs action,
whereas the second term is a higher derivative (HD) action.

  (HD actions were recently discussed by Jansen, Kuti and Liu~\cite{JKL}. 
Here we are interested in a different critical point from the
one studied in \rf{JKL}. At the technical level, this allows us
to introduced only a laplacian-squared HD term, whereas for the purpose
of \rf{JKL} it was crucial to introduce also a laplacian-cubed one.)

  The action \seeq{s1} is gauge invariant, where the lattice gauge 
transformation is given by $U_{x,\m}\to g_x U_{x,\m} g^\dagger_{x+\hat\m}$
and $\f_x\to g_x \f_x$ for $g_x\in G$. Now, since $\f_x\in G$ too, we may
use the lattice gauge invariance to completely eliminate the $\f_x$
field. Note that this operation affects only $S_H$.
We introduce the notation
\beq
  S_V(U) =  S_H(\f,U) \Big|_{\f_x=I} \,.
\label{sv}
\eeq
The subscript of $S_V$ stands for ``vector''.  
$S_H$ can be recovered from $S_V$ by making
the substitution $U_{x,\m}\to \f_x^\dagger U_{x,\m} \f_{x+\hat\m}$.
(The significance of the $\f_x$ field, 
which is associated with  the longitudinal degrees of freedom,
is discussed in Sect.~\ref{sect5} and Sect.~\ref{sect6}.)

  We will denote the first formulation of the theory (\seeq{s1}) as
the {\it Higgs picture}. The alternative formulation (\seeq{sv}) where
only $U_{x,\m}$ (but not $\f_x$) is present, 
is called the {\it vector picture}.
The equality of the partition functions in the two pictures
extends to observables. Any observable in the vector picture is
mapped to a gauge invariant observable in Higgs picture, and vice versa.
Thus, we are dealing with two mathematically equivalent formulations
of the same theory~\cite{phi}. 

  In this section we assume $\tk \gg 1$. The physics in this parameter range
is more easily accounted for in the vector picture, 
which is used below to study the classical potential, 
and to set up the weak-coupling expansion.
In the vector picture, the gauge non-invariance of the 
action  resides in $S_V$. Therefore, we will ultimately demand that
the marginal terms in $S_V$ have the form of a gauge fixing action,
{\it cf.}\seneq{gcond}.

\subsection{The classical potential}
\label{sect2c}

For simplicity we consider the classical potential
in the U(1) case. The essential features generalize to the non-abelian case.
Making use of the standard weak-coupling expansion
\beq
  U_{x,\m} = \exp(ig_0 A_{x+{\hat\m\over 2},\m}) \,,
\label{A}
\eeq
and considering a constant $A_\m$ field, 
the action $S_V$ leads to the following classical potential
\beq
  V_{\rm cl} = \k\, \cf(g_0 A_\m) + \tk\, \cf^2(g_0 A_\m) \,,
\label{vcl}
\eeq
\beq
  \cf(g_0 A_\m) = 2\sum_\m \Big( 1-\cos(g_0 A_\m) \Big) \,.
\label{cosA}
\eeq
Note that $S_G$ is zero for a constant abelian field.

  For $\k>0$, the absolute minimum of the classical
potential is $A_\m=0$ (mod $2\p/g_0$). 
Since the quadratic term in $V_{\rm cl}$ comes only
from the $\k$-term, a non-zero vector condensate arises for $\k<0$.
The classical features of the transition can be determined by
keeping only the leading term in the expansion of $\cf(g_0 A_\m)$,
separately for the $\k$- and $\tk$-terms. (This approximation
is consistent for $|\k| \ll \tk$.) The result is the
quartic potential
\beq
  V_{\rm cl} \approx \k g_0^2\,\sum_\m A_\m^2 
               + \tk g_0^4\,\Big( \sum_\m A_\m^2 \Big)^2 \,.
\label{vclx}
\eeq
\seEq{vclx} closely resembles the potential
of a $\F^4$ theory. For small negative $\k$, the minimum is
\beq
   \Big|\hspace{-.3mm}\Big|\svev{A_\m}\Big|\hspace{-.3mm}\Big| =
   {1\over g_0} \left( {|\k| \over 2\tk} \right)^\half \,,
   \quad\quad \k<0 \,.
\label{vevA}
\eeq
\seEq{vevA} exhibits the mean-field critical exponent $1/2$.
It is easy to check that this is the absolute
minimum of the classical potential. 

  We note that \seeq{vevA} is invariant under arbitrary SO(4) rotations, 
reflecting the symmetry of the approximate potential \seeq{vclx}. 
When higher-order corrections 
are taken into account, the rotational symmetry of the potential 
is reduced to the lattice hypercubic symmetry.

  Below, the phase with a non-zero vector condensate will be denoted as
the FMD phase. We will speak about the FMD transition, referring to
the transition from the rotationally invariant phase to the 
FMD phase in the large-$\tk$ region.
FMD stands for {\it ferromagnetic directional}. The preferred spacetime 
direction of the FMD phase is defined by the vectorial VEV. 
For $g_0 \ne 0$, there are no Goldstone bosons in the 
FMD phase, because the lattice rotation group is discrete.
The limiting $g_0 = 0$ theory is discussed in Sect.~\ref{sect5}, 
and in particular
we explain there in what sense the FMD phase is ferromagnetic.
 
\subsection{The weak-coupling expansion}
\label{sect2d}

We now want to study fluctuations around the classical vacuum
$U_{x,\m}=I$ (equivalently $A_\m=0$) in the rotationally-invariant phase,
close to the FMD transition where the theory defined 
by eqs.~(2-5) is expected to be critical. The FMD transition
is given by $\k=0$ in the classical approximation. As mentioned earlier,
we are assuming $\tk \gg 1$. We therefore focus on the HD term 
in its vector picture form.
Relaxing the assumption of a constant $A_\m$ field, we find
\bqry
  & & \left. \tk\, \f^\dagger\bo^2(U)\f  \right|_{\f_x=1} = \NON
  & & \tk g_0^2 \left( \Big( \sum_\m \D^-_\m A_\m \Big)^2
  + g_0^2 \Big( \sum_\m A_\m^2 \Big)^2 +\cdots \right) \,.
\label{hda}
\eqry
where the dots stand for irrelevant operators. $\D^-_\m$ is the backward
lattice derivative, defined as $\D^-_\m f_x = f_x - f_{x-\hat\m}$ for
any function $f_x$.

  \seEq{hda} contains a longitudinal kinetic term. We define 
\beq
  {1\over 2\x_0} \equiv \tk g_0^2 \,,
\label{a0}
\eeq 
and we will assume that $\x_0$ is an $O(1)$ parameter. 
This means that the longitudinal kinetic term
belongs to the tree-level lagrangian. Remember that a transversal 
kinetic term is provided by the gauge invariant plaquette action $S_G$. 
Finally, we assume that the tree-level vector boson mass is zero.
Under these assumptions, the tree-level vector propagator is
\beq
  G_{\m\n}(p) = { \P^\perp_{\m\n}(\hat{p})
           + \x_0 \P^\parallel_{\m\n}(\hat{p}) 
           \over \hat{p}^2 } \,,
\label{vprop}
\eeq
where
\bqry
  \P^\perp_{\m\n}(\hat{p}) & = & \d_{\m\n}-\hat{p}_\m\hat{p}_\n/\hat{p}^2 \,,\\
  \P^\parallel_{\m\n}(\hat{p}) & = & \hat{p}_\m \hat{p}_\n/\hat{p}^2 \,,
\eqry
and $\hat{p}_\m = 2\sin(p_\m/2)$. 

  Massless weak-coupling perturbation theory is defined by the vector 
propagator \seeq{vprop}, and by a set of vertices which can be read off from
the lattice action using \seeq{A} in the usual way.

\subsection{The gauge-fixing action}
\label{sect2e}
 
In view of the presence of kinetic terms for all polarizations,
lattice perturbation theory is manifestly {\it renormalizable}.
According to the standard lore,
renormalizability implies a Lorentz invariant {\it scaling} behaviour 
in the vicinity of the gaussian critical point $g_0=1/\tk=0$.
The scaling behaviour is achieved by tuning a finite number of counter-terms,
that correspond to the relevant and the marginal operators.

  At this stage, the marginal gauge symmetry breaking terms in the tree-level 
vector action (see \seeq{hda}) do not have the form of a gauge-fixing action,
{\it cf.}\seneq{gcond}. The way to remedy this is to
add another term to the HD action. There are two options.
The new term can be chosen to cancel the quartic term in \seeq{hda}.
The remaining marginal term -- the longitudinal kinetic term -- 
has the form of a gauge-fixing action for the linear Lorentz gauge 
$\partial\cdot A =0$.
Alternatively, the new HD term can lead to a mixed marginal term
proportional to $(\partial\cdot A) A^2$. In this case one recovers the
non-linear gauge $\partial\cdot A + g A^2 = 0$.

  The linear gauge $\partial\cdot A =0$ is more familiar,
and less complicated to implement in perturbation theory.
Moreover, the above non-linear gauge is consistent only for U(1) 
or SU(N)$\times$U(1), whereas the linear gauge is consistent for 
any gauge group. The linear gauge has, however, one technical
disadvantage. The quartic term in \seeq{hda} is the stabilizing
term of the classical potential (see \seeq{vclx}). In its absence,
one has to reanalyze the classical potential, and make sure that
it is stabilized by a higher-power term (in practice this is an $A^6$ term).
This task is done in \rf{gfx}, which is henceforth referred to as $II$.

  Here we will consider only the non-linear gauge. Since the necessary
mixed term contains a derivative, one can modify the HD action while
leaving the classical potential intact. This simplifies our task, as
the large-$\tk$ study of the phase diagram in Sect.~\ref{sect2c} remains valid. 
The new HD action is
\beq
  S_{HD}^{\rm n.l.} = {1\over 2 \x_0 g_0^2}\, \tr \sum \left(
           \f^\dagger \bo^2(U) \f + 2B \sum_\m \D^-_\m V_\m  \right) \,,
\label{shd}
\eeq
\beq
  B_x = \sum_\m \left( {V_{x-\hat\m,\m} + V_{x,\m} \over 2} \right)^2 \,,
\label{B}
\eeq
\beq
   V_{x,\m} = {1\over 2i}
   \left(\f^\dagger_x U_{x,\m} \f_{x+\hat\m} - \hc \right) \,.
\label{V}
\eeq
Going to the vector picture and applying the weak-coupling expansion,
we have 
\beq
  V_\m \big|_{\f_x=I} = g_0 A_\m - {1\over 6} (g_0 A_\m)^3  + \cdots \,.
\label{vexp}
\eeq 
It is easy to check that the desired mixed term is now present 
in the tree-level vector action. We comment that, in the Higgs picture, 
$V_\m$ is a gauge-invariant local vector field, whose
expectation value serves as an order parameter for the FMD phase.
(The corresponding order parameter in the vector picture 
is the expectation value of $V_\m \big|_{\f_x=I}$.) At the classical level,
$\svev{V_\m} = g_0 \svev{A_\m}$, where the latter is given by \seeq{vevA}.

  For the laplacian-squared HD action (see \seeq{sh}), it is evident that 
$U_{x,\m}=I$ is the unique absolute minimum for {\it all} configurations,
and not only for the constant ones considered in the
classical potential (we assume $\f_x=I$). This property, which is necessary to 
validate the weak-coupling expansion, applies to the new HD 
action\seneq{shd} as well. The proof is given in $II$.
The symmetric combination used in the definition of $B_x$ (\seeq{B}),
which does not affect the marginal term contained in $B \sum_\m \D^-_\m V_\m$,
is essential for the proof.

  We define the lattice gauge-fixing action to be
\beq
  S_{\rm gf}^{\rm n.l.}(U) \equiv S_{HD}^{\rm n.l.}(\f,U) \Big|_{\f_x=I} \,. 
\label{shdsgf}
\eeq
As expected, $S_{\rm gf}^{\rm n.l.}$
has the classical continuum limit  $(1/2\x_0)(\partial\cdot A + g A^2)^2$.
Because of the irrelevant terms it contains,
one cannot write $S_{\rm gf}^{\rm n.l.}$ as the (sum over $x$ of the) 
square of a local function of the $U_\m$-s. Consequently,
the gauge-fixed lattice action is not invariant under BRST transformations.

  It is interesting that the breaking of
(gauge and) BRST invariance is a common feature of the chiral fermion action
and the gauge-fixing action. In the case of the gauge-fixing action,
it has to be so because of a theorem by Neuberger~\cite{hn},
which asserts that any lattice BRST-invariant (gauge-fixed) partition function 
must vanish due to lattice artefact Gribov copies. 
We return to the role of lattice Gribov copies in Sect.~\ref{sect4}.

  Before we introduce fermions, the complete lattice action
(in the vector picture) is therefore
\beq
  S_V^{\rm n.l.} = S_G + S_{\rm gf}^{\rm n.l.} + S_{\rm fp}^{\rm n.l.} 
  + S_{\rm ct} \,.
\label{sfull}
\eeq
For the non-linear gauge, the continuum Faddeev-Popov action involves the 
operator (we suppress the group structure constants) 
$\partial^2 + ig A\cdot\partial + g \{ A,\partial+igA \}$. 
The last term is absent in the case of the linear gauge.
For the discretization of $\partial^2$
we take the standard (free) lattice laplacian. We are discretizing a 
second order operator, and our choice avoids the 
appearance of any FP doublers. For the interaction terms, any lattice operator
with the correct classical continuum limit should do.
For the discretization of $g A\cdot\partial$, for example,
one can take $\sum_\m V_\m \D_\m$ where $\D_\m$ is the antisymmetric
difference operator. (Since BRST symmetry is broken anyway by the gauge-fixing 
action, we make no attempt to preserve any exact relation between the
discretized versions of $\partial^2$ and $A\cdot\partial$.) We note that
the ghost fields contribute to the effective potential only through loops,
and so they do not modify the tree-level considerations.

$S_{\rm ct}$ is the counter-term action. 
The role of $S_{\rm ct}$ is to enforce BRST invariance
in the low momentum limit of lattice perturbation theory~\cite{roma}.
The BRST symmetry is violated in particular by (marginal)
SO(4)-breaking lattice operators. Therefore, enforcing BRST invariance should 
also restore full SO(4)-invariance in the continuum limit. 
The counter-term action is more naturally 
written in terms of $A_\m$. We define $S_{\rm ct}$ as a local
functional of the $U_\m$-s by trading $A_\m$ with $V_\m$ using \seeq{vexp}. 
(The second term in the expansion of $V_\m$, which breaks SO(4) invariance, 
is needed only for the dimension-two mass counter-term.
In all other cases one simply replaces $g_0 A_\m$ with $V_\m$.)

  The BRST identity $m_r^2=0$, which says that 
the (renormalized) vector boson mass must vanish to all orders in 
perturbation theory, is consistent with taking the continuum limit at 
the FMD transition. As a mass counter-term one can take the $\k$-term 
in \seeq{sh}. This means that $\k$ is tuned to $\k_{c.l.}(g_0,\x_0)$, where
in perturbation theory $\k_{c.l.} = \sum_{n\ge 1} c_n(\x_0)
g_0^{2(n-1)}$. Note that the coefficient of the mass term in
\seeq{vclx} is $\k g_0^2$. The absence of an $O(1/g_0^2)$ term in 
the expansion of $\k_{c.l.}$ is in agreement
with the vanishing of the tree-level vector boson mass. 

  In this paper we have simplified things by considering only the most
important counter-term, namely, the dimension-two mass term.
In the case of the non-linear gauge, the next most important counter-term
is the dimension-four SO(4)-breaking term $\sum_\m A_\m^4$.
As for the linear gauge, the classical potential is stabilized by
an $A^6$ term, and $(\sum_\m A_\m^2)^2$ too occurs only as a counter-term. 
In this case, the effect of the dimension-four non-derivative 
counter-terms is discussed in $II$. One finds that the
conventional (hyper-cubic invariant) phase and the FMD phase both
extend into the higher-dimensional phase diagram. Also, the FMD 
transition remains continuous when the dimension-four counter-terms
are tuned to their critical values. The crucial features leading
to these conclusions are
(a) it is justified to expand $U_\m$ up to a finite order in $A_\m$
(equal to the dimension of the stabilizing term)
when looking for the  absolute minimum of the potential;
(b) the coefficients of the counter-terms are $O(1)$, whereas the
coefficient of the gauge-fixing action is $O(1/g_0^2)$.
Since these features are true in the case of the non-linear gauge as well,
we expect a similar robustness against the inclusion of additional
counter-terms.

  In this section we have discussed the phase diagram only in the large-$\tk$
limit. The phase diagram for arbitrary $\k$ and $\tk$ is studied in
Sect.~\ref{sect5a}. This study, as well as additional arguments presented
in Sect.~\ref{sect6a}, further clarify why the continuum limit 
of the gauge-fixing approach should be defined at the FMD transition.

\section{Chiral fermions}
\label{sect3} 

  In a gauge invariant lattice theory, the minimum of the plaquette action 
is unique up to a gauge transformation, and the transversal kinetic term
is sufficient to define a valid weak-coupling expansion. In the {\it absence} 
of gauge invariance, there exists a valid weak-coupling expansion 
provided the gauge-fixing action of Sect.~\ref{sect2e} 
is added to the plaquette action.
This applies also to the gauge-fixing action presented in $II$ for
the linear gauge $\partial\cdot A = 0$. Using either of these gauge-fixing
actions, there is a lot of freedom in the choice of the chiral fermion action.
We consider here (in the vector picture) an
action which is the most economic in the number of fermionic degrees of
freedom~\cite{roma2}. (For related work see \rf{zrgz,pryor}.) 
Other fermion actions have certain advantages over the one presented here,
and in particular they can reduce the required fine-tuning.

  According to \rf{roma2}, one introduces a two-component lattice
fermion field $\c_x$, to account for a single Weyl fermion in the 
target continuum theory. The fermion action is (suppressing coordinates
summations)
\beq
  S_F = \sum_\m \bar\c\, \s_\m D_\m(U) \,\c 
        - {w\over 4} ( \c \bo \c + \hc) \,,
\label{sf}
\eeq
\beq
  D_{xy,\m}(U) = {1\over 2}( \d_{x+\hat\m,y} U_{x,\m} 
  - \d_{x-\hat\m,y} U^\dagger_{y,\m}) \,.
\label{dmu}
\eeq
Here $\bo_{xy}$ is the free lattice laplacian (\seeq{bo} for $U_{x,\m}=I$),
and $\c_x \c_y \equiv \e_{\a\b}\c_{x,\a} \c_{y,\b} = \c_x^T \e \c_y$ 
where $\e$ is the antisymmetric two-by-two matrix. We assume $w=O(1)$.
The first term in \seeq{sf} is the naive lattice discretization of the
continuum Weyl action. The second term is a Majorana-Wilson (MW) term,
that breaks explicitly gauge invariance as well as the fermion number
symmetry. The latter is unwanted, because fermion number is not
conserved in the continuum theory.

  In order to understand the properties of the lattice fermion path integral,
it is convenient to recast the fermion action \seeq{sf} in terms of
four components fields $\j_M$ and $\bar\j_M$. By definition,
$P_L \j_M = \c$ and $P_R \j_M = \e \bar\c^T$, where
$P_{R,L}=\half(1 \pm \g_5)$ denote chirality projectors. 
$\bar\j_M$ is not an independent field, and is given by 
\beq
  \bar\j_M \equiv \j^T_M C \,.
\label{maj}
\eeq
Here $C$ is the antisymmetric four-by-four charge conjugation matrix,
obeying $C^2=-1$, $\g_\m^T C = - C \g_\m$ and $\g_5 C = C \g_5$.
In terms of these four component fields, the fermion action takes the form
\bqry
  S_F & = & {1\over 2} \sum_\m \bar\j_M \Big[ 
        \g_\m D_\m(U) P_L + \g_\m D_\m(U^*) P_R \Big] \j_M \NON
      & &  - {w \over 4}\, \bar\j_M \bo \j_M \,.
\label{smaj}
\eqry

Let us first examine \seeq{smaj} in perturbation theory.
The tree-level fermion propagator is the (massless) 
Wilson propagator~\cite{roma2}.
Because of the Majorana-like condition \seeq{maj}, 
the symmetry factors in Feynman graphs are the same as for Majorana fermions.
Now, if we go to the small momentum limit, we find that the two chiralities
of $\j_M$ couple to the gauge field according to {\it complex conjugate} 
representations of the gauge group. One sees that the role of \seeq{maj} 
is to consistently maintain the identification $P_L \j_M \leftrightarrow \c$, 
$P_R \j_M \leftrightarrow \bar\c$, at the level of Feynman diagrams.
That \seeq{smaj} correctly describes a single left-handed Weyl fermion,
can be verified by calculating the non-analytic part
of one-fermion-loop diagrams,
that should agree with the continuum result in the limit of
a vanishing external momentum (the role of counter-terms is discussed below). 
As usual, the non-analytic contribution comes 
from an infinitesimal neighbourhood of the origin in the Brillouin zone.
In this neighbourhood one can neglect  the Wilson term in both numerators 
and denominators. The left-handed and right-handed components of $\j_M$
are no longer coupled, and one can reexpress the Feynman integrand 
in terms of the continuum propagator for a single Weyl fermion.
(A left-handed fermion loop is equal to 
a right-handed fermion loop in the complex conjugate representation.
The ``double-counting'' is compensated by a one-half symmetry factor 
for each closed fermion loop, which arises from the 
Majorana-like condition \seeq{maj}.)

We next discuss the rigorous definition of the fermionic path integral.
We introduce the $2N\times 2N$ fermion matrix $Q_{\a\b}$ by writing 
the action in the generic form 
$S_F = {1\over 2} \sum_{\a,\b} \j_M^\a\, Q_{\a\b} \,\j_M^\b$. 
(The charge conjugation
matrix $C$ is absorbed into the definition of $Q$.) It is easy to check that
$Q$ is antisymmetric. The fermion path integral takes the following 
form~\cite{maj}
\bqry
& & \int \prod_\a d\j_M^\a\, \exp\left( {1\over 2} 
  \sum_{\a,\b} \j_M^\a\, Q_{\a\b} \,\j_M^\b \right) \NON
& &  = {1\over 2^N N!}\, \e_{\a_1,\b_1,\ldots,\a_N,\b_N}\,
  Q_{\a_1,\b_1} \cdots Q_{\a_N,\b_N} \NON
& &  \equiv  {\rm pf}\left({Q\over 2}\right)
\label{pf}
\eqry
There is no integration over the (dependent) variables $\bar\j_M$.
According to \seeq{pf}, the fermionic path integral is a {\it Pfaffian}.
In general, ${\rm pf}(Q/2)$ is complex,
as expected from the euclidean path integral for a single Weyl fermion.

(As a further check that our fermion path integral describes a Weyl
fermion, we can consider a
``two-generation'' model, where each complex representation occurs twice
in the fermion spectrum. Using the identity 
${\rm pf}^2(Q/2)= \det(Q)$, that holds for
a general antisymmeric matrix, this two-generation model can be
defined by an action similar to \seeq{smaj}, where we now
drop an overall one-half factor, substitute
$\j_M \to \j_D$, $\bar\j_M \to \bar\j_D$, and regard $\j_D$ and $\bar\j_D$
as {\it independent} Dirac-like variables. The counting of degrees of
freedom is now straightforward. Since the two chiralities of $\j_D$
belong to complex conjugate representations, this action actually describes 
two left-handed Weyl fermions in the {\it same} complex representation.)

Until now we have implicitly discussed the fermions in the background
of a fixed external gauge field. The main result of the previous section 
is that, with the gauge-fixing action, perturbation theory 
is valid for a dynamical gauge field as well. 
Therefore, with appropriate counter-terms, 
{\it the continuum fields describing the scaling 
behaviour are in one-to-one correspondence with the massless poles of the 
various tree-level propagators}. 
If we choose an anomaly-free fermion spectrum,  the 
scaling region should be governed by a continuum chiral gauge theory,
in the relevant gauge. 
We note that if one chooses an {\it anomalous} fermion spectrum,
the scaling region will still be governed by a renormalizable
lagrangian, but BRST invariance and, hence, 
unitarity will be violated.

Finally, let us discuss the fermion mass counter-terms.
As with ordinary Wilson fermions, a mass counter-term
$m_0\, \bar\j_M \j_M$ is necessary to maintain the masslessness of each 
chiral fermion. (A different fermion action that does not require
mass counter-terms will be discussed elsewhere~\cite{BGS2}.)
The renormalized Majorana-like mass is proportional to $(m_0-m_c)$,
where $m_c$ is (minus) the fluctuations-induced mass.
If $(m_0-m_c)$ is small but non-zero, BRST invariance will be 
explicitly broken in the scaling region.
The scaling behaviour is then governed by a renormalizable continuum theory
which is not gauge invariant (hence also non-unitary).
By tuning $m_0$ to $m_c$, assuming all other counter-terms
already have their critical values, we recover BRST invariance
simultaneously with the masslessness of the chiral fermions.
(The situation on the lattice is similar 
to what one would encounter in the continuum, if a gauge non-invariant
regularization is employed for a chiral gauge theory. 
As on the lattice, Majorana-like mass 
counter-terms may be needed, alongside with other
gauge non-invariant counter-terms, to cancel the breaking of gauge invariance
induced by the regularization, and to ensure that 
the renormalized amplitudes are gauge invariant.)

  An important question in the literature on lattice chiral gauge theories,
is how to correctly reproduce fermion number violation in the continuum
limit. Different solutions have been 
proposed to the problem~\cite{EP,DM,BD,BHS}. 
We hope that the present approach can shed new light on it.

  For definiteness, we adopt the strategy of \rf{roma2}.
Namely, we demand that the lattice fermion action should have no
symmetry which is not present in the target continuum theory. Now, while 
the action \seeq{sf} is not invariant under global U(1) transformations
with an arbitrary phase, it is still invariant under the residual
{\it discrete} symmetry $\c \to -\c$, $\bar\c \to - \bar\c$.
This symmetry implies a (mod 2) conservation law for each fermion species, 
which still causes a problem. Consider for definiteness an SU(5) GUT, 
with one generation that contains a $\overline{\bf 5}$ and a ${\bf 10}$. 
In an instanton background, the numbers of zero modes for these 
representations are respectively one and three. 
This is in conflict with the above (mod 2) conservation laws.
Thus, on top of the MW terms present in \seeq{sf} 
for each representation, one has to introduce an additional gauge-noninvariant 
MW term that couples the $\overline{\bf 5}$ and the ${\bf 10}$.
(As a result, a Majorana-like mass counter-term that mixes 
the $\overline{\bf 5}$ and the ${\bf 10}$ will be necessary  too.)
With this new MW term, the remaining discrete symmetry leads only   
to (mod 2) conservation of the total fermion number for each generation.

\section{Why gauge fixing}
\label{sect4}

  We now return to the lattice effective action,
and consider some of its properties in more detail. 
We keep the discussion at an informal level.
Our approach has been presented in detail in the previous sections,
and the aim here is to clarify the nature of the {\it problems} 
that it is meant to solve. 

  In this section we assume that the lattice chiral fermion action $S_F$
is bilinear in the fermion fields, and that it depends in addition
only on the link variables $U_{x,\m}$. (This corresponds to the 
vector picture.) It is also assumed that $S_F$ 
is (mildly) local. The lattice spacing $a$ will be shown explicitly
in this section.

  The difficulties encountered in the construction of lattice chiral 
gauge theories can be addressed at a more rigorous level~\cite{ys,rev}.
This complementary discussion, which focuses on
the robustness of the Nielsen-Ninomiya theorem, is given in Sect.~\ref{sect6}.

\subsection{Rough lattice gauge transformations\protect\\ and the need
for gauge fixing}
\label{sect4a}

The lattice effective action is defined by integrating 
out the fermions
\beq
   S_{\rm eff}(U) = - \log \int \cd\j\cd\bar\j\, e^{-S_F(U, \j, \bar\j)} \,.
\label{seff}
\eeq
Clearly, the well-defined object is $\exp(-S_{\rm eff})$, rather than
$S_{\rm eff}$ itself. For our purpose it will be sufficient to
consider the {\it perturbative} effective action, and so we will
ignore the problems associated with the global definition of $S_{\rm eff}$.

  The variation of $S_{\rm eff}$, in response to an infinitesimal 
lattice gauge transformation at the point $x_0$, has
the following general form
\beq
   \d_{x_0} S_{\rm eff} \approx
   c_0\, \co_{x_0}^{con} + \sum_{n\ge 1} a^n \sum_i c_{n,i} \co_{x_0}^{n,i}
\label{an}
\eeq
The $\approx$ sign indicates that the \rhs is computed
perturbatively. Note that the gauge field 
is external, and so the gauge-field action is not needed at this stage.
$\co_x^{con}$ and $\co_x^{n,i}$ are local lattice operators that depend on
$A_\mu$, {\it cf.} \seeq{A}. The dimension of $\co_x^{n,i}$ is $4+n$,
and the $i$-summation is over linearly independent operators
of this dimension. $\co_x^{con}$ is some discretized version 
of the consistent anomaly. We assume that all operators of
dimension less than or equal to four, other than $\co_x^{con}$, 
have been cancelled by counter-terms. If, moreover, 
we choose a set of complex representations that satisfies the anomaly 
cancellation condition, then $c_0=0$.

  The infinite sum on the \rhs of \seeq{an} accounts for
lattice artifact violations of gauge invariance. The precise
form of these violations is model dependent, but their existence
is generic. As can be easily seen by going to momentum space, 
which is the usual setting for a perturbative computation, 
this sum represents a double expansion in $|g_0 a A_\m|$ and $|a p_\m|$. 

  Let us first consider a smooth gauge field $A_\m^0$ which is characterized
by some physical scale $\L_{\rm phys} \ll a^{-1}$, and the corresponding 
configuration of link variables $U_{x,\m}^0$ defined via \seeq{A}. 
Since the dimensionful quantities $A_\m$ and $p_\m$ are $O(\L_{\rm phys})$,
both of the above expansion parameters are small.
\seEq{an} is the gradient of $S_{\rm eff}$ with respect to a motion
inside the lattice gauge orbit. Therefore, $S_{\rm eff}$ is (approximately)
constant on the orbit in the vicinity $U_{x,\m}^0$.
The constancy of $S_{\rm eff}$ extends to that portion of the 
orbit which is reachable from $U_{x,\m}^0$ by a smooth
gauge transformation.

  The problem is that, on the lattice, smooth gauge transformations 
represent a tiny part of the local gauge group. 
Let $U^{(g)}_{x,\m} = g_x U_{x,\m}^0 g_{x+\hat\m}^\dagger$ be a {\it generic}
configuration in the orbit of $U_{x,\m}^0$, and let $A_\m^{(g)}$
be related to $U^{(g)}_{x,\m}$ via \seeq{A}. 
Since the operators that occur on the \rhs of \seeq{an}
are {\it not} gauge-invariant, they are sensitive to the value of $g_x$
for $x$ in the vicinity of $x_0$.
Now, the $g_x$-s on different sites a {\it uncorrelated}. 
As a result, the expansion parameters $|g_0 a A^{(g)}_\m|$ 
and $|a \D_\n A^{(g)}_\m / A^{(g)}_\m|$ are $O(1)$ for a generic 
lattice gauge transformation.  
We conclude that $S_{\rm eff}$ fails to be (approximately) constant
on most of the lattice gauge orbit. This is true for {\it any} orbit, 
including orbits that have a smooth representative.

\subsection{Proliferation of lattice Gribov copies}
\label{sect4b}

The above problem stems from the 
{\it roughness} of generic lattice gauge transformations.
Following \rf{roma} we make no attempt to reduce the violations of 
gauge invariance at the level of the effective action. 
Instead, our aim is to suppress the Boltzmann weight
of rough gauge field configurations  
(relative to smooth configurations that belong to the same orbit) 
consistently with the gauge invariance of the physical Hilbert space,
namely, via {\it gauge fixing}. 

  In lattice QCD, gauge fixing is a matter of choice, since it has no
effect on the gauge-invariant observables. Here, the fermion action is not
gauge invariant. As a result, the gauge-fixing method is an 
{\it integral part} of the definition of the theory.
Different gauge-fixing methods may in general give rise to different
phase diagrams with different critical points. There is no guarantee
that every gauge-fixing method will lead to a non-trivial continuum limit,
let alone to a chiral gauge theory.

  Still, in order to make progress, one has to choose {\it some}
gauge-fixing method. Vink~\cite{vink} proposed to use the laplacian gauge,
where a maximally-smooth representative is chosen on each
gauge orbit by global minimization. The laplacian gauge is highly non-local, 
and this creates difficulties both in the analytic and in the numerical study
of this method. 

  As discussed in detail in Sect.~\ref{sect2}, we build a local
gauge-fixing lattice action, that (a) has the unique absolute
minimum $U_{x,\m}=I$, and (b) reduces to a covariant gauge-fixing
action in the classical continuum limit. 
Our gauge-fixing actions (eqs.~(16-18) for the non-linear gauge,
see $II$ for the linear gauge) are clearly not the most naive
discretizations of the corresponding continuum actions.
Focusing for simplicity on the linear gauge $\partial\cdot A = 0$, 
let us examine what goes wrong with a naively-discretized gauge-fixing action.
We thus consider the following action
\beq
  S_{\rm gf}^{\rm naive} = {1\over 2\x_0 g_0^2}\, \tr \sum_x \cg_x^2 \,,
\label{snaive}
\eeq
\beq
  \cg_x =  \sum_\m \D^-_\m V_{x,\m}
        = {1\over 2i} \sum_\m \big( \D^-_\m\, U_{x,\m} - \hc \Big) \,.
\label{Fx}
\eeq
Note that $g_0^{-1} \sum_\m \D^-_\m V_{x,\m}$ reduces in the classical 
continuum limit to $\partial \cdot A$, as it should.

  What is common to our gauge-fixing action(s) and to the naive gauge-fixing
action \seeq{snaive}, is that they contain a longitudinal 
kinetic term. The trouble with $S_{\rm gf}^{\rm naive}$ is that it supports
a {\it dense set} of  Gribov copies for the identity field $U_{x,\m}=I$. 
Each of these Gribov copies is a classical vacuum of 
$S_G+S_{\rm gf}^{\rm naive}$.
The superposition of contributions coming from all these classical vacua,
which {\it cannot} be calculated perturbatively, 
may ultimately render the fermion spectrum vector-like.
Even without fermions, stability of the classical potential is lost,
and it is unclear whether the FM-FMD transition (associated with
a divergent vector-field correlation length) can be maintained in the
weak-coupling limit. (If the gauge-fixing action in \seeq{sfull}
is replaced by $S_{\rm gf}^{\rm naive}$, the resulting classical potential is
identically zero. If we assume $\k=O(1/g_0^2)$, which implies the presence
of a tree-level mass term, one has 
$V_{\rm cl}^{\rm naive} = \k\, \cf(g_0 A_\m)$ (compare \seeq{vcl}). 
The minimum of $V_{\rm cl}^{\rm naive}$ is $A_\m =0$ for $\k>0$, and 
$A_\m =\p/g_0$ for $\k<0$.)

  We now demonstrate the existence of a dense set of lattice 
Gribov copies for the identity field. Consider first the U(1) case. 
The condition $\cg_x = 0$ is satisfied if the imaginary part of $U_{x,\m}$ 
is zero everywhere. The latter
is true if we consider only lattice gauge transformations where $g_x=\pm 1$.
In other words, in spite of the presence of the gauge-fixing action 
$S_{\rm gf}^{\rm naive}$, the Gribov copies of the identity field still
exhibit a local Z$_2$ symmetry. (This is also true for the non-linear 
gauge, if one replaces $\cg_x$ in \seeq{snaive} by 
$\cg_x^{\rm n.l.} =  \sum_\m ( \D^-_\m V_{x,\m} + V_{x,\m}^2)$.)
An ``elementary'' Gribov copy is created if we choose $g_x= -1$
for $x=x_0$, and $g_x=1$ elsewhere. This clearly shows
that the Gribov copies are {\it local lattice artefacts}.

  The above example generalizes to non-abelian groups.
In the case of SU(2N) and SO(2N) groups, simply replace $\pm 1$ by $\pm I$.
Moreover, for any SU(N) group,
one can choose an SU(2) subgroup (which for simplicity we assume to
lie at the top left corner) and repeat the above construction
with $g_x=diag(\pm 1, \pm 1, 1, 1, \ldots)$.
The discrete local symmetry of the Gribov copies is therefore larger
than Z$_2$ in the general case. 
For a more detailed discussion of lattice Gribov copies see \rf{copies}.

  A number of remarkable similarities draw us to say that the
{\it proliferation of Gribov copies} is the spin-1 counter-part of 
the {\it fermion doubling} problem. In both cases,
one deals with the discretization of a first-order differential
operator: in the spin-$\half$ case, this is the Dirac (or Weyl) equation;
in the spin-1 case, this is a covariant gauge condition 
(see Sect.~\ref{sect2e}).
In both cases the problem arises when a
non-compact continuum variable is replaced with a compact lattice
variable: fermion doubling arises because, unlike in the continuum, 
the lattice momentum is periodic; in the spin-1 case, 
also the non-compact continuum gauge field
is replaced with compact group variables.
In both cases, there are theorems that establish an impasse
under certain mild-looking conditions: 
the Karsten-Smit~\cite{KS} and Nielsen-Ninomiya~\cite{NN} theorems
which predict fermion doubling, and Neuberger's theorem~\cite{hn} which 
asserts that any BRST-invariant partition function must vanish identically.
Finally, in both cases the solution is to reduce the symmetry 
of the lattice theory, by adding {\it irrelevant} terms
to the naively-discretized action. In the case of Wilson fermions, 
this is the role of the Wilson term. As for our gauge-fixing
action(s), one can show (see $II$) that 
it is equal to the square of a discretized gauge condition,
plus irrelevant terms. Thus, again, the irrelevant terms reduce the symmetry,
this time by breaking explicitly BRST invariance.

\subsection{Other approaches}
\label{sect4c}

A different approach to the dynamical problems created by
rough lattice gauge transformations is to adopt a more
sophisticated definition for the effective action. 
The prominent representatives of this approach are the 
interpolation method~\cite{intp,twolat} and the overlap formalism~\cite{ovlp}. 

  In the interpolation method, one constructs a continuum interpolating field
$A_\m^{\rm cont} = A_\m^{\rm cont}(y;U_{x,\m})$, $y \in R^4$,
for each configuration of the lattice gauge field.
The determinant of the Weyl operator,
$\s \cdot (\partial +ig A^{\rm cont})$, is then defined using
a separate regulator. Associated with the fermion regularization is a new
cutoff parameter, denoted generically $\L_{\rm f}$, 
which must be sent to infinity before the lattice
spacing $a$ is sent to zero. (A concrete method~\cite{twolat} is to 
discretize the Weyl action on a finer lattice with a lattice spacing 
$a_{\rm f} \ll a$, using {\it e.g.} the fermion action of \rf{roma2} 
(\seeq{sf}). In this case $\L_{\rm f} = a_{\rm f}^{-1}$.)

  Consistent regularizations of the Weyl determinant
break gauge invariance for finite values of the cutoff,
even when the fermion spectrum is anomaly-free.
Therefore, gauge-noninvariant counter-terms are needed in the 
interpolation method too. (It has been proposed that gauge-noninvariant 
counter-terms may be avoided, if the real part of the effective action 
is regulated separately from the imaginary part~\cite{sep}. 
In spite of attempts in this direction~\cite{twolat}, 
it remains unclear whether this procedure can be implemented beyond
perturbation theory without violating locality, and, eventually, unitarity.)

  For given $U_{x,\m}$-s, the interpolating field assigns a
{\it local winding number} to each hypercube (in the non-abelian case), 
or to each plaquette (in the abelian case). In the non-abelian case,
this is the winding number of the continuum gauge transformation
defined on the faces of the hypercube, that brings the interpolating
field to a prescribed axial gauge; in the abelian case, the continuum
gauge transformation is defined on the perimeter of each plaquette.
Now, a fundamental requirement is that the fermion 
determinant should be gauge invariant in the limit $\L_{\rm f}\to \infty$. 
Gauge invariance can be established only if 
$A_\m^{\rm cont}(y)$ is globally bounded~\cite{intp,twolat}.
Gauge-invariance is therefore recovered in the limit $\L_{\rm f}\to \infty$
only on that portion of the lattice gauge orbit, where all the local winding 
numbers are zero~\cite{rev}.
The solution is to apply a gauge transformation that sets all local winding
numbers to zero {\it before} computing the fermion determinant
(for simplicity we consider a trivial global topology).
We note that the smoothing gauge transformation is non-local,
and so a careful study of potential problems associated with the
infinite-volume limit is required.

  In the overlap approach, while the real part
of the effective action is gauge invariant by construction,
the imaginary part is not.  Again, we expect that gauge-noninvariant 
counter-terms will be needed, starting at some finite loop order. 
Potentially severe problems with the overlap approach 
were pointed out in \rf{modwg}. According to our judgement,
subsequent works (including in particular \rf{ovlp2}) 
fail to address the issues raised in \rf{modwg}.
Numerical evidence for the lack of gauge invariance 
(in the non-abelian case) has been found in \rf{KM} (see fig.~1 therein).

\section{The reduced model}
\label{sect5}

  Returning to our approach, we consider in this section the limit of a
vanishing gauge coupling. Since $1/g_0^2$ is the coefficient of the 
plaquette action, the $g_0=0$ limit constrains the lattice gauge field 
to the trivial orbit. 

  The theory defined by $g_0=0$ limit is called the {\it reduced model}. 
If we use the {\it vector picture}, the reduced model is 
obtained by substituting $U_{x,\m} \to \f_x^\dagger \f_{x+\hat\m}$
in the lattice action. The lattice gauge field measure $\prod\int d U_{x,\m}$ 
is replaced by $\prod\int d \f_x$. (Alternatively, starting from the
{\it Higgs picture} that already involves both $U_{x,\m}$ and $\f_x$ 
(see Sect.~\ref{sect2b}), 
one obtains the reduced model by simply setting $U_{x,\m}=I$.)

  In the weak gauge-coupling limit, the transversal modes are perturbative 
at the lattice scale. Many important features, including the fermion
spectrum, are determined by the dynamics of the longitudinal modes.
The utility of the reduced model is that it allows us 
to study the longitudinal dynamics in isolation, without making any a-priori
assumption. The reduced model accounts for dynamical situations 
ranging from a divergent longitudinal correlation length,
as in our approach, down to a very short correlation length.
In this section we study a prototype reduced model. The entire
dynamical range is realized in different regions of its phase diagram.
In Sect.~\ref{sect6} we discuss the effects of the longitudinal dynamics on
the fermion spectrum, first in general terms and then in our approach.

\subsection{The phase diagram}
\label{sect5a}

In this subsection (and the next one) we discuss the reduced model 
related to the lattice action of Sect.~\ref{sect2b}, for $\f_x \in {\rm U}(1)$.
Since \seeq{sh} is written in the Higgs picture, the reduced
model is obtained by setting $U_{x,\m}=1$. This leads to the action
\beq
  S'_H = \sum \left(
         -\k\, \f^\dagger\bo\f + \tk\, \f^\dagger \bo^2 \f \right)\,.
\label{sh'}
\eeq
Our first task is to derive the mean-field phase diagram in the
$(\tk,\k)$-plane. This phase diagram is in fact generic, and pertains also
to the more relevant theories defined in Sect.~\ref{sect2e} and in $II$.

  Let us first consider the ordinary VEV, $v$, as an order parameter
(or the staggered VEV $v_\AM$). 
An additional order parameter will be introduced shortly. By definition
\bqry
  v & = & \svev{\f_x} \,, 
\label{vev} \\
  v_\AM & = & \svev{\e_x \f_x}\,,
\label{vevam}
\eqry
where $\e_x=(-1)^{\sum_\m x_\m}$. 
For $\tk=0$, we recover the familiar non-linear sigma model.
On the $\k$-axis there is a symmetric (PM) phase for $|\k| < \k_c$, 
a ferromagnetic (FM) phase for $\k > \k_c$, and an
antiferromagnetic (AM) phase for $\k < -\k_c$.
The field redefinition $\f_x\to \e_x \f_x$ maps $\k$ to $-\k$,
thus implying a symmetry of the $\k$-axis.

  We now extend the discussion to the full $(\tk,\k)$-plane.
Mean-field approximation in $d$-dimensions yields the 
following equation for the FM-PM line
\beq
  \k + (2d+1)\tk = \k_c \,,\quad\quad \mbox{FM-PM}.
\label{crt}
\eeq
The equation for the AM-PM line is
\beq
  \k + (6d-1)\tk = -\k_c \,,\quad\quad \mbox{AM-PM}.
\label{crta}
\eeq
The FM-PM and AM-PM transitions are continuous.
The symmetry of the $\k$-axis extends to $\tk \ne 0$.
Under the field redefinition $\f_x\to \e_x \f_x$, the point $(\tk,\k)$
is mapped in four dimensions to $(\tk,-\k-32\tk)$. 
This implies that the linear equation
\beq
  \k + 16\tk = 0 \,,
\label{sym}
\eeq
defines a symmetry line of the phase diagram. The FM-PM and AM-PM 
lines meet in the second quadrant, at the point
$(-{\k_c\over 7}, {16\k_c\over 7})$ on the symmetry line.
It can be shown that, beyond this point, the symmetry line
is a first-order transition line separating the FM and AM phases~\cite{BGS1}.

  In condensed matter physics, it is well-known that spin models with
competing interactions tend to develop
a ground state that breaks translation and rotation invariance.
If a small antiferromagnetic interaction is added to a dominant 
ferromagnetic one, the spin orientation of the ground state
will {\it rotate} slowly with a wave vector $q_\m\ne 0$ 
(see \rf{cm} for a recent review).

  In the reduced model defined by \seeq{sh'}, competing interactions 
occur when $\k$ and $\tk$ have opposite signs. In order to look for
a similar phenomenon, we introduce the mean-field ansatz
\beq
  \svev{\f_x} = v\, e^{i q_\m x_\m} \,.
\label{hlc}
\eeq
We assume $0 \le q_\m < 2\p$.
A non-zero $q_\m$ signals the spontaneous breaking of translation
and rotation invariance. More precisely, translation invariance is broken in 
the direction defined by $q_\m$, but it remains unbroken in the transversal 
directions. In condensed matter, phases with a non-zero $q_\m$ are known as
{\it helicoidal-ferromagnetic} ones. Here, the 
helicoidal-ferromagnetic phase of the reduced model is the $g_0=0$ boundary
of the FMD phase of the full theory (see Sect.~\ref{sect2c}), and the name FMD 
will be used both in the full theory and in the reduced model. 

  A simple mean-field method is based on a  
{\it factorized probability measure} (see {\it e.g.} \rf{P}).
For the FM-PM transition, one can use the following  
factorized probability measure
\beq
  \cp_0(\th_x) = { 1 + 2v\cos(\th_x) \over 2\p} \,.
\label{prob}
\eeq
In order to accommodate a non-zero $q_\m$, we generalize this to
\beq
  \cp(\th_x) = { 1 + 2v\cos(\th_x - q_\m x_\m) \over 2\p} \,.
\label{probx}
\eeq
One has $\svev{1}_\cp = 1$ and
$\svev{ e^{i\th_x} }_\cp = v\, e^{i q_\m x_\m}$ 
(in agreement with \seeq{hlc}). Because of its factorized nature, 
the $q_\m$-dependence of $\cp(\th_x)$ affects only 
the {\it internal energy}, but not the {\it entropy}.
Introducing the notation $S'_H = \sum_x \ch_x$, we find
\beq
  \svev{\ch}_\cp = (1-v^2) (8\k + 72\tk)
                   + v^2 \left( \k\, \cf(q_\m) + \tk\, \cf^2(q_\m)
  \right) \,,
\label{vclq}
\eeq
where the function $\cf$ is defined in \seeq{cosA}.
While the choice of the factorized probability measures \seeqs{prob} 
and\seneq{probx} is somewhat arbitrary, the internal energy\seneq{vclq} 
is a universal feature of any mean-field approximation for $S'_H$.

  If we consider a point in the phase diagram well to the right of
the FM-PM line, the value of $v$ is finite in lattice units. 
Making the self-consistent assumption that $q_\m$ is small, 
the location of the FM-FMD transition can be determined by minimizing the
internal energy with respect to $q_\m$. Remarkably, the $q_\m$-dependent 
part of the internal energy coincides with the
classical potential\seneq{vcl}, if we make the identification 
$g_0 A_\m \leftrightarrow q_\m$. 
Consequently, there is complete agreement between the mean-field properties of
the FM-FMD transition in the reduced model, and the classical 
properties of the FMD transition in the $g_0 \ne 0$ theory. 
The mean-field location of the FM-FMD transition is $\k=0$.
For $\k>0$ one is in the FM phase, whereas for $\k<0$ one is in the
FMD phase. Close to the FM-FMD line, $q_\m$ is given by 
\seeq{vevA} where $g_0 A_\m$ is replaced by $q_\m$ (after this
replacement $g_0$ drops out, and one has $q^2 = -\k/2\tk$ for a small 
negative $\k$). The FM-FMD line ends when it hits the FM-PM line.
The multi-critical point where the PM, FM and FMD phases meet is known
as a Lifshitz point~\cite{lp}. Its mean-field value is $({\k_c\over 9},0)$.
(Lifshitz points exhibit rich critical behaviour. This was discussed 
recently in a field theoretic context by J.\ Kuti~\cite{jk}.)

  The remaining features of the mean-field phase diagram are 
as follows (see FIG.~1).
The transformation $\f_x\to \e_x \f_x$ maps $q_\m$ to
$q_\m + \p$. This implies the existence of a second FMD region 
(and another Lifshitz point) below the symmetry line in the fourth quadrant. 
The PM-FMD line, separating the paramagnetic phase from the FMD phase,
can be determined by first minimizing 
the (internal) energy with respect to $q_\m$,
and then the (free) energy with respect to $v$.
The PM phase occupies a bounded region in the phase diagram. 
The two FMD regions above and below the symmetry line belong to a 
single FMD phase. The PM-FMD line lies on an ellipse, with the FM-PM 
and AM-PM lines tangent to it at the two Lifshitz points. A more 
detailed mean-field calculation will be presented elsewhere~\cite{BGS1}. 

  The mean-field ansatz\seneq{hlc} implies the
simultaneous breaking of the internal U(1) symmetry, as well as of
rotation and translation invariance. We believe
that one cannot break translation invariance without at the 
same time breaking an internal symmetry.
However, one can conceive of a phase (denoted PMD) where
only rotation symmetry is broken, while the internal symmetries as well as 
translations are unbroken. The order parameter for a PMD phase of the
reduced model is the expectation value of the
composite vector field (compare \seeq{V})
\beq
  V_{x,\m}^\parallel 
  =  {1\over 2i} \left(\f^\dagger_x \f_{x+\hat\m} - \hc \right) \,.
\label{Vpar}
\eeq
At the moment, however, we have no evidence for a PMD phase.

  The phases of the reduced model are depicted in TABLE~I. 
The order parameter $v_H$ is defined as
$v_H = \svev{\f_x\, e^{-i q_\m x_\m}}$.  Note that in the
special case $q_\mu=(0,0,0,0)$ ($q_\mu=(\p,\p,\p,\p)$),
$v_H$ coincides with $v$ ($v_\AM$).
In this paper we usually do not distinguish between $v$ and $v_H$, 
since the correct meaning can be understood from the context.
However, this distinction is important
in numerical simulation. The value of $q_\m$, to be used in
the measurement of $v_H$, can be determined for example by
measuring $\svev{\f^\dagger_x \f_{x+\hat\m}}$ and extracting its phase.

  The relation between the $(\tk,\k)$-phase diagram of the reduced model and 
the $(\tk,\k,g_0)$-phase diagram of the full theory is the following. 
In the U(1) case, the symmetric (PM) phase is the boundary of a
Coulomb phase, and the broken (FM or AM) phase is the boundary of a 
Higgs phase. In the non-abelian case, the PM, FM and AM phases correspond
to the boundary of a single Higgs-confinement phase.
Finally, the helicoidal-ferromagnetic (FMD) phase of the reduced model 
is the boundary of the FMD phase of the full theory.

  In the $g_0 \to 0$ limit of the full theory, we find (approximately) 
massless gauge bosons close to the FM-FMD line, 
as well as close to the PM-FM line,
and in the entire PM phase. An interesting observation is that,
even without fermions, if we want to study a Lorentz gauge-fixed Yang-Mills
theory on the lattice, then the appropriate critical line is the FM-FMD line.
The reason is that, in order to keep the longitudinal kinetic term in
the tree-level action, $\tk$ must scale like $1/g_0^2$ (see Sect.~\ref{sect2}). 
In the large-$\tk$ limit, a PM phase does not
exist, and criticality can only be 
achieved by approaching the FM-FMD line.

\subsection{The weak-coupling expansion\protect\\ in the reduced model}
\label{sect5b}

In view of the properties of the weak-coupling expansion in the 
full theory (Sect.~\ref{sect2}), 
and in particular \seeq{a0}, one expects that $1/\tk$
will play the role of a coupling constant in the reduced model.
We will now demonstrate this explicitly. We do not carry out here
any detailed calculations that require the full lattice Feynman rules.
Therefore, we work in the continuum approximation, {\it i.e.}
we extract from the lattice action the marginal and relevant
terms, that control the critical behaviour in the vicinity of the
gaussian critical point $1/\tk=0$.

  The weak-coupling expansion is facilitated by expanding around a
broken symmetry vacuum. We first introduce the Goldstone boson (GB) field
$\th_x$ via $\f_x=e^{i\th_x}$. (This classical expansion is consistent
with the mean-field ansatz\seneq{hlc}, because $v \to 1$ for $\tk \to \infty$.)
Rescaling $\th \to (1/\sqrt{2\tk})\,\th$ we find, in the continuum 
approximation, the following GB lagrangian
\beq
  \cl_{GB} = {1\over 2} \int d^4x\, 
             \left(  (\bo\th)^2 
             + {\k\over\tk}\, \partial_\m\th\, \partial_\m\th
             + {1\over 2\tk}\, (\partial_\m\th\, \partial_\m\th)^2 \right) \,.
\label{gb}
\eeq
\seEq{gb} is valid on the FM side of the transition line. 
On the FMD side, one first looks for
the classical vacuum by assuming $\th = q_\m x_\m$ and minimizing
for $q_\m$. The result is the same as in the mean-field approximation.
The weak-coupling expansion on the FMD side is then defined via
$\th \to q_\m x_\m + (1/\sqrt{2\tk})\,\th$.

  We will consider here only the FM side.
Taking the Fourier transform of the bilinear part of the lagrangian,
we find the following  GB propagator
\beq
  G_0^{-1}(p) = (p^2)^2 + m_0^2\, p^2 \,,
\label{gb0}
\eeq
where $m_0^2=\k/\tk$. Analytical continuation to
Minkowski space shows the existence of a positive-residue pole at $p^2_M=0$,
and a negative-residue (ghost) pole at $p^2_M=m_0^2$. 
These poles merge into a quartic singularity in the limit $m_0^2 \to 0$.
The continuum limit of the reduced model is therefore not unitary. 
(In Sect.~\ref{sect6b} we discuss the interaction of the GB field with 
fermions. The crucial requirement is that the non-unitary GB sector,
which accounts for the two unphysical polarizations of the gauge bosons,
will decouple from the fermions in the continuum limit.
This decoupling is discussed in detail in \rf{BGS2}.)

  Because of the quartic kinetic term in the GB lagrangian, 
the canonical dimension of the GB field $\th(x)$ is zero. 
The GB lagrangian is invariant under the shift symmetry
$\th(x) \to \th(x)+const$, which forbids the appearance of 
non-derivative terms under renormalization.
In addition, the GB lagrangian is invariant under
the discrete symmetry $\th(x) \to -\th(x)$. 
\seEq{gb} is the most general
renormalizable lagrangian allowed by these symmetries.

  What marks the Feynman rules of the GB lagrangian, is that
one derivative acts on every line attached to a vertex. 
The derivatives acting on the two ends of each internal line effectively cancel
one factor of $1/p^2$ in the propagator. The result is that
the UV power counting of the GB model is
the same as in an ordinary $\l\F^4$ theory. If we ignore
the vector index carried by the partial derivatives, one can match each
term in the GB lagrangian with a corresponding term in the
$\l\F^4$ lagrangian according to the rule $\partial\th\to\F$.

  In the $\l\F^4$ theory, at the one loop level only
the mass term is renormalized, but not the kinetic term.
By analogy, in the GB lagrangian only 
$(\partial_\m\th)^2$, but not $(\bo\th)^2$, is renormalized at 
the one-loop level. The induced one-loop $(\partial_\m\th)^2$-term 
has a positive sign. This implies  
\beq
  \k_{c.l.}(\tk) =  -c + O\left( \tk^{-1} \right)\,.
\label{kcl}
\eeq
(Note that the coefficient of $(\partial_\m\th)^2$ in
\seeq{gb} is $\k/\tk$.)
The dimension of the positive constant $c$ is two. Its numerical
value, which is $O(1/a^2)$, has to be determined by a lattice calculation.
Finally, as in the $\l\F^4$ theory, the one-loop beta-function 
is determined by the vertex renormalization, and is found to be
positive. Explicitly
\beq
  \b(\tk^{-1}) \equiv \L {\partial\over\partial\L}\, \tk^{-1}
  = {5\over 16\p^2}\, \tk^{-2} \,,
\eeq
where $\L$ is the UV cutoff (the inverse lattice spacing 
in a lattice calculation).

\subsection{Infra-Red divergences of the critical theory}
\label{sect5c}

If we tune $\k$ to $\k_{c.l.}(\tk)$, the quadratic kinetic term
in \seeq{gb0} vanishes, and the renormalized GB propagator reads
\beq
  G_r^{-1}(p) = Z\, (p^2)^2 \,,
\label{gbr}
\eeq
where $Z$ accounts for the wave-fuction renormalization.
This quartic propagator leads to IR divergences in four dimensions,
like massless bosons with an ordinary kinetic term do in two dimensions.

  The IR divergences of massless Goldstone bosons lead to the restoration 
of continuous symmetries in two dimensions~\cite{nossb}. 
Only symmetric observables, which are invariant under all
the continuous symmetries, can have a non-zero value. 
There are theorems~\cite{el,d,dd} that
guarantee the IR-finiteness of the symmetric observables.

  Here the quartic propagator\seneq{gbr}
does not characterize a whole phase, but only the FM-FMD line itself. 
The order parameter $v_H$ (or $v$) dips close to
the FM-FMD line, and vanishes on that line in several (may be in all)
interesting cases~\cite{BGS1,BGS2}. 
The theorems on the finiteness of symmetric observables, in particular \rf{dd},
generalize to four dimensions. Also, as in two dimensions, the
predictions of the weak-coupling expansion are often valid,
if interpreted carefully~\cite{ew}. 
This will be important in Sect.~\ref{sect6b}.

\subsection{Realistic reduced models}
\label{sect5d}

The key features of the simplified reduced model studied in
this section extend to the realistic reduced models,
defined from the (gauge-fixing and ghost) action of Sect.~\ref{sect2e} for the 
non-linear gauge, or the action of $II$ for the linear gauge. 
This includes the qualitative structure of the phase diagram
and in particular the FM and FMD phases,
the gaussian critical point at $\tk=\infty$, 
and the IR divergences on the FM-FMD line.

  Particularly interesting are the critical FM-FMD theories 
in the reduced models that correspond to the linear gauge 
$\partial\cdot A=0$. In the abelian case, 
the linear-gauge reduced model leads to a free theory with a $1/(p^2)^2$
propagator. The properties of this critical theory
are analogous to the spin-wave phase of a two-dimensional abelian theory.
This will be discussed in detail elsewhere~\cite{BGS1,BGS2}.
The critical theory for a non-abelian gauge group
was investigated by Hata~\cite{hata} in the continuum approximation.  
His main result is that, like non-abelian sigma models in two dimensions, 
these four-dimensional non-linear models are asymptotically-free.
It will be interesting to investigate the significance of this
result for the construction of gauge-fixed non-abelian lattice
theories via our approach.

\section{Fermions in the reduced model}
\label{sect6}

  In a manifestly gauge invariant theory like QCD, 
the fermion spectrum can be read off from the lattice action by going to 
the free field limit $g_0=0$. Here, the fermion action is
not gauge invariant (in the vector picture), and the limit $g_0=0$ gives rise
to an interacting theory, namely, to the reduced model. 
We identify the elementary fermions of a general lattice gauge theory 
with the independent {\it fermionic massless poles} of  
the associated reduced model. (If the fermion action is gauge invariant,
any $\f_x$-dependence of its reduced-model form can be eliminated by
a field redefinition.) It is justified
to determine the matter spectrum by setting $g_0=0$, since,
in a scaling region, the transversal degrees of freedom are perturbative 
at the lattice scale.

\subsection{The robustness of the No-Go theorems}
\label{sect6a}
 
Let the gauge field belong to a Lie group $G$. 
By construction, the associated reduced model has 
a global $G$-symmetry, denoted $G_L$, that acts on $\f_x$ by 
left multiplication. (The reduced model is obtained from
the vector picture via $U_{x,\m} \to \f_x^\dagger \f_{x+\hat\m}$, and the 
product $\f_x^\dagger \f_{x+\hat\m}$ is invariant under left multiplication. 
Notice also that the gauge-invariant Higgs picture can be obtained 
by gauging the $G_L$ symmetry of the reduced model.)
Now, we demand the existence of massless vector bosons 
in the scaling region, which can be identified with the gauge bosons 
of the target continuum theory. These vector bosons couple to the 
Noether current associated with the $G_L$ symmetry. 
Thus, assigning the fermions to representations of $G_L$
determines whether the continuum limit is chiral or vector-like.

  In previous chiral fermion proposals, it was usually attempted to take 
the continuum limit in a symmetric phase, where the $G_L$ symmetry is
{\it not} broken spontaneously. (Physical gauge invariance 
is restored dynamically in a symmetric phase, when we consider the 
full $g_0 \ne 0$ theory. This means that there are no light
unphysical states, whose decoupling in the continuum limit requires
fine-tuning. Since the VEV of the $\f_x$ field is zero, the physics 
in a symmetric phase is more easily accounted for in the gauge-invariant 
Higgs picture.) In a symmetric phase, the fluctuations 
of the $\f_x$ field are usually not controlled by any small parameter. 
As a result, non-perturbative methods had to be
invoked in order to determine the fermion spectrum. Where available, 
it was always found that the true fermion spectrum is vector-like
(see \rf{rev,revDP,pmsfer} for details).

  We have discussed this phenomenon in \rf{ys,rev}, and argued that
it has a simple physical explanation. Here we can only outline the 
key considerations leading to this conclusion, and we refer
the reader to \rf{ys,rev} for the details. One starts with the observation
that, in a symmetric phase of the reduced model, 
there are generically no massless scalars. 
Therefore, the only massless particles (if any) are fermions.
(It could be~\cite{pmsfer} that no massless fermions are present
unless a mass term if fine-tuned. Since we are in symmetric phase,
a massless fermion obtained by fine-tuning is necessarily a Dirac
fermion.) Now, in four dimensions, there are no renormalizable interactions 
involving only fermion fields. The continuum limit defined by a
generic point inside a symmetric phase is therefore
a theory of free massless fermions (if it is not empty).
One can then construct an {\it effective lattice 
hamiltonian} for the fermions, that satisfies all 
the assumptions of the {\it Nielsen-Ninomiya theorem}. 
(The effective hamiltonian is defined as
the $p_0 = 0$ limit of the inverse of a suitable two-point function.) 
We refer here in particular to the analytic structure near the zeros of 
the effective hamiltonian, and to the existence of a smooth interpolation 
throughout the rest of the Brillouin zone. 
This leads to the  conclusion that 
the fermion spectrum is vector-like in a symmetric phase,
provided the underlying theory is local.
(In the case of a non-local theory one expects violations of unitarity
and/or Lorentz invariance, see \rf{rev} for references to the original
literature.)

  This impasse extends, by continuity, to the fermion spectrum 
on any phase transition line that separates a symmetric phase from a
broken phase. In particular, even though the gauge boson mass vanishes
on the PM-FM line, we do not expect to find a chiral gauge theory 
by taking the continuum limit at the PM-FM line.
The fermion spectrum will be vector-like if the PM-FM line is approached
from the PM phase. If we approach the PM-FM line
from the FM phase, we can only obtain a mirror fermion model~\cite{revIM},
but we cannot decouple the unwanted mirror fermions.

\subsection{Evading the No-Go theorems}
\label{sect6b}

Let us now investigate what changes when the continuum limit
is taken at the FM-FMD line.
We will consider the simplest case, namely a U(1) gauge group 
with the gauge-fixing action pertaining to the linear gauge, {\it cf.} $II$. 
As mentioned in Sect.~\ref{sect5d}, 
the properties of the critical FM-FMD theory
are analogous to the spin-wave phase of a two-dimensional abelian theory.

  We go from (the vector picture of) the full 
theory to the reduced model according to the rule 
$U_{x,\m} \to \f_x^\dagger \f_{x+\hat\m}$.
The fermion action \seeq{sf} becomes (we use the two-component notation)
\beq
  S'_F = \sum_\m (\bar\c \f^\dagger) \s_\m \D_\m (\f\c) 
        - {w\over 4} \Big( \c \bo \c + \hc \Big) \,.
\label{sf'}
\eeq
The fermion variables in \seeq{sf'} are {\it neutral} with 
respect to $G_L$. If, instead, we use the {\it charged} variables
$\c_c = \f\c$, the fermion action reads
\beq
  S'_F =  \sum_\m \bar\c_c\, \s_\m \D_\m \c_c 
  - {w\over 4} \Big( (\f^\dagger\c_c) \bo (\f^\dagger\c_c) + \hc \Big) \,.
\label{sfc}
\eeq

  According to the rules of the weak-coupling expansion 
(see Sect.~\ref{sect5b}),
the tree-level fermion action is obtained by substituting the classical
vacuum $\f_x=1$. Using \seeq{sfc} for definiteness, we get
\beq
  S^0_F = \sum_\m \bar\c_c\, \s_\m \D_\m \c_c 
        - {w\over 4} \Big( \c_c \bo \c_c + \hc \Big) \,,
\label{sf0}
\eeq
In the limit $w=0$, only the kinetic term 
$\sum_\m \bar\c_c\, \s_\m \D_\m \c_c$ is left. Thus, the $w=0$ action
exhibits the infamous doubling, with sixteen Weyl fermions altogether.
(Each fermion is associated with a point in the Brillouin zone,
whose lattice momentum components are equal to either 0 or $\p$.)
Since we take $w=O(1)$, the MW term eliminates the doublers, and the pole in
the tree-level fermion propagator describes a single Weyl field 
(see Sect.~\ref{sect3}). 

  Had we started from the fermion action
written in terms of the neutral variables (\seeq{sf'}),
the substitution $\f_x=1$ would lead to a tree-level action identical to
\seeq{sf0}, but with the neutral field $\c$ replacing the charged 
field $\c_c$. Now, deep in the FM phase this makes no difference, 
because the $G_L$ symmetry is broken anyway by the $\f_x$-VEV, which is $O(1)$
in lattice units. However, the $G_L$ symmetry
is restored right on the FM-FMD line~\cite{BGS1,BGS2}. 
It is therefore a meaningful (and important) question to ask what
are the $G_L$-quantum numbers of the massless fermions.

  The fact that one cannot simply read off the quantum numbers
of one-fermion states from the tree-level action,
is a consequence of the IR-divergent
nature of the GB propagator, $1/(p^2)^2$. The way to proceed is to examine 
a family of fermionic two-point functions  
$\G_n=\svev{(\f^n\c)\, (\bar\c \f^{\dagger n})}$. 
Assuming all mass parameters have been tuned to their critical values,
$\G_n$ will in general contain terms proportional to 
$(\sl{p})^{-1} \log^k(p^2)$ for any $k$.
The presence of logarithmic terms (which typically lead to power law
corrections when summed over all orders) means that the operator $\f^n\c$ 
{\it does not create a one-particle state}~\cite{ew,pmsfer}. 
Only when there are no logarithmic corrections do we have a simple
massless pole, and the quantum numbers of the intermediate one-fermion state
must coincide with the quantum numbers of the interpolating fermion field.

  The fermion spectrum in the reduced model can
be studied in detail using the weak-coupling expansion.
While the actual calculations require a substantial amount of work,
the conclusions are robust, as they really depend on universal
properties of the low energy effective (continuum) lagrangian. 
A one-loop calculation, which is also supported by
numerical simulations, will be presented elsewhere~\cite{BGS2}.
Here we will list the key results, as they apply to the MW fermion action.

\begin{itemize}

\item Logarithmic terms are {\it absent} only for $n=1$,
namely in the two-point function $\svev{\c_c\, \bar\c_c}$. 
Consequently, the massless fermion has the 
quantum numbers of the $\c_c$ field. The latter is {\it charged},
which means that the $\c_c$ Weyl fermion will couple to the transversal 
gauge field, when the latter is turned on.

\item As can be expected on general grounds (see Sect.~\ref{sect3}),
divergent Majorana-like mass terms are induced at the one-loop
level. These must be cancelled by suitable counter-terms,
to maintain the masslessness of each chiral fermion.
 
\item When the Majorana-like mass counter-terms are tuned to their critical
values, the unphysical GB field {\it decouples} 
from the $\c_c$-fermions. The continuum limit is a direct product of
(in general) several free theories, one associated with 
the unphysical GB field, and one associated with every species of
chiral $\c_c$-fermions.

\end{itemize}

\noindent 
This establishes an agreement between the predictions of the weak-coupling
expansion in the full theory and in the reduced model,
thus supporting the consistency of our approach.
The properties of the reduced model are true for an arbitrary 
fermion spectrum, and this is consistent with the vanishing of the
anomaly in the absence of a transversal gauge field. 

  In comparison with previous chiral fermion proposals (see Sect.~\ref{sect6a})
we note two key differences that allow us to escape from a similar impasse. 
First, one may worry that the need to tune mass counter-terms may indicate
that we got the wrong spectrum (\eg Dirac instead of Weyl fermions).
Now, when the (Majorana-like) mass counter-terms are {\it not} tuned to their
critical values, the fermions remain {\it coupled} to the IR-singular GB field
in the low-energy limit. Due to potential IR divergences, 
it is not at all clear that (massive)
one-fermion states could be consistently defined in this case,
nor that such states would have well-defined $G_L$-quantum numbers.
The off-critical theory remains to be investigated in the future.
However, in view of the above IR subtleties,
the general conclusion is that by considering the role of 
fermion mass perturbations, one does not end up with an argument {\it against} 
the existence of a chiral spectrum at the critical point.
(See also the discussion of fermion mass counter-terms in Sect.~\ref{sect3}.)

  The other key difference is that the
continuum limit is now taken at the phase transition separating
{\it two broken phases} of the reduced model. 
Off the FM-FMD line (on both sides) the $G_L$ symmetry is broken 
spontaneously, and all asymptotic states do not have well-defined
$G_L$-quantum numbers.  
On the FM-FMD line itself, the $G_L$ symmetry is restored, and
the question arises whether we do not run into the same old conflict with
the No-Go theorems. The answer is contained in the analytic structure
discussed above. Thanks to the presence of
the highly IR-singular GB field, a zero in the inverse propagator 
does not necessarily imply the existence of a one-fermion state with
the same quantum numbers. As an illustration, consider the four-component
unit-charge field $\j_c$, whose left-handed component is $P_L \j_c = \c_c$, 
and whose right-handed component is $P_R \j_c = \f^2 \bar\c_c$. 
If we consider the inverse two-point function of $\j_c$, we may erroneously
conclude that it interpolates a massless Dirac fermion.
In reality, only the left-handed channel of this inverse propagator
has a simple zero $\sim$ $\sl{p}$, implying the existence of
a unit-charge left-handed fermion. In the right-handed channel,
one finds a $\sl{p}\log(p^2)$ correction in the one-loop approximation,
which implies the absence of a right-handed fermion with the same
charge.

\section{Open questions} 
\label{sect7}

  In Sect.~\ref{sect2e}, the criterion for fixing the counter-terms was 
to enforce BRST invariance (and, hence, unitarity) order by order. 
This perturbative prescription is incomplete. 
Ultimately, the counter-terms should be determined by a
non-perturbative method. To rigorously define the continuum limit, one has to
specify a trajectory in the Higgs (or Higgs-confinement) phase, 
that ends at the gaussian point $g_0=1/\tk=0$ on the FMD boundary. 
(See Sect.~\ref{sect5a} for the phase diagram.) 
In addition, one has to construct
a BRST operator, and prove its nilpotency in the continuum limit.
Enforcing BRST invariance should also lead to the restoration of full SO(4)
invariance, because the marginal SO(4)-breaking
operators violate the BRST symmetry too.

  We comment that similar problems are encountered in lattice QCD
with Wilson fermions, where the axial-flavour symmetries are broken on
the lattice, in analogy with the BRST symmetry in our gauge-fixing approach.
When using Wilson fermions, tuning is required not only at the level of 
the lattice action, but also in the construction of renormalized operators 
with well defined axial-flavour transformation properties~\cite{axial}.
This is analogous to the problem of defining BRST-invariant
operators in our gauge-fixing approach. (In QCD, the fine-tuning problem
can be solved using domain-wall fermions~\cite{dk,FS,BS}.
Whether a similar solution exists for the tuning problem in
the gauge-fixing approach, is an interesting question.)

  Our gauge-fixing formulation can be tested by applying it to
asymptotically-free gauge theories which are {\it not} chiral.
In particular, in the absence of fermions,
one should study whether the confining behaviour and the mass gap of 
Yang-Mills theories are reproduced. One possibility is that the 
FMD transition becomes weakly first-order due to non-perturbative effects. 
This scenario is favourable, at least from the point of view
of numerical simulations. Another possibility is that the correlation length
of the vector field strictly diverges at the FMD transition, already 
for $g_0 \ne 0$. In this case, a consistent
continuum limit may exist provided all the massless excitations
are unphysical.

\section{Conclusions}
\label{sect8}

  In a {\it regularized} chiral gauge theory, the longitudinal modes of 
the gauge field couple to the fermions. Before the regularization is 
removed, there are violations of gauge invariance 
even if the fermion spectrum is anomaly-free. 
When we use the lattice regularization, the longitudinal modes should decouple 
in the continuum limit, but it may be too much to expect for exact
decoupling when the lattice spacing is still finite.

  The gauge-fixing approach aims to decouple the longitudinal modes
in the continuum limit. In this paper we have discussed how the gauge-fixing
approach may be realized, thus making the first step of a
systematic investigation of the gauge fixing approach.
We have constructed a lattice gauge-fixing action 
that has a unique classical vacuum. The gauge-fixing action
contains a longitudinal kinetic term, and leads to a 
renormalizable weak-coupling expansion,
which is valid even if the lattice fermion action is not gauge invariant. 
We have argued that the continuum fields, needed to describe the 
scaling behaviour, are in one-to-one correspondence with the poles 
of the tree-level lattice propagators. This should accommodate
any consistent theory, including anomaly-free chiral gauge theories. 

\acknowledgements

  This work evolved out of an attempt to examine the feasibility, 
as well as the consequences, of tuning the gauge bosons mass to zero deep 
in the broken phase. The necessity for an FMD phase was 
clarified to me during a discussion with Moshe Schwartz. 
Questions and comments of Karl Jansen and Maarten Golterman were vital 
for the subsequent development of this work. Its present form I owe
to an on-going collaboration with Wolfgang Bock and Maarten Golterman.
I also thank Ben Svetitsky and Yoshio Kikukawa for their comments 
on an earlier version of this paper.





\begin{figure}  
\centerline{
\epsfysize=9.0cm
\epsfbox{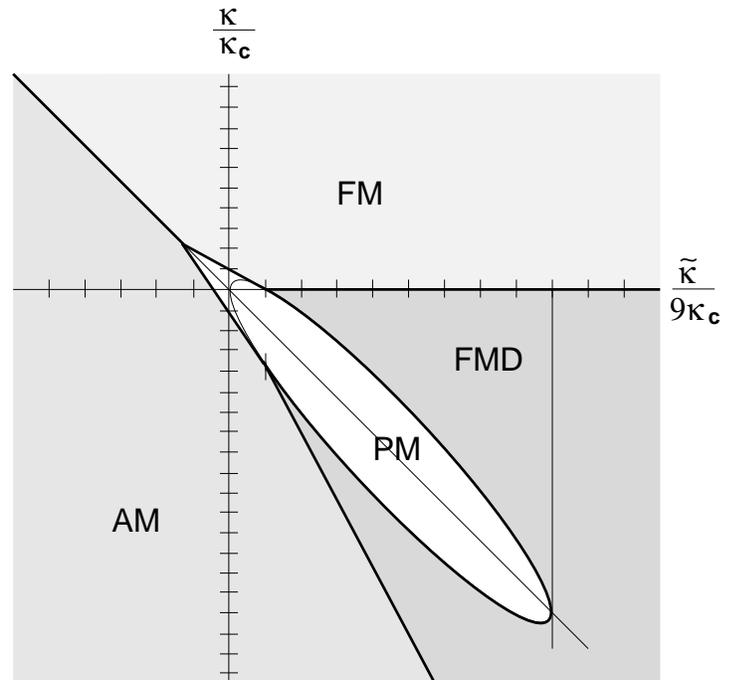}                         
}
\caption{ \noindent Mean-field phase diagram. See TABLE I for the 
definition of the various phases.}
\label{phase_diagram}
\end{figure}

%
\begin{table}
\caption{
Phases of the reduced model. 
The entries indicate which order parameters are non-zero in each phase.
}
\begin{tabular}{|c|c c c c|}     
\hline
phase & $v$ & $v_\AM$ & $v_H$ & $V^\parallel_\m$
\\ \hline \hline
PM & no & no & no & no
\\ \hline
FM & yes & no & yes & no
\\ \hline
AM & no & yes & yes & no
\\ \hline
FMD & no & no & yes & yes
\\ \hline
PMD(?) & no & no & no & yes
\\ \hline
\end{tabular}
\end{table}

\end{document}